# Extension of the Glauber-Sudarshan Mapping For Classical and Quantum Energy Spectra

Paul J. Werbos[1]
ECCS Division, National Science Foundation

## Abstract

This paper begins by reviewing the general form of the Glauber-Sudarshan P mapping, a cornerstone of coherence theory in quantum optics, which defines a two-way mapping between ensembles of states S of any classical Hamiltonian field theory and a subset of the allowed density matrices ρ in the corresponding canonical bosonic quantum field theory (QFT). It has been proven that Tr(Hρ)=E(S), where E is the classical energy and H is the normal form Hamiltonian. The new result of this paper is that ensembles of S of definite energy E map into density matrices which are mixes of eigenstates of eigenvalue zero of N(H-E,H-E) where N represents the normal product. This raises interesting questions and opportunities for future research, including questions about the relation between canonical QFT and QFT in the Feynman path formulation.

## 1. Introduction and Summary

This work was motivated by two practical questions: (1) to what extent can simulations of classical partial differential equations (PDE), such as those used in nuclear phenomenology [1], approximate the correct predictions of energy spectra from the corresponding bosonic quantum field theory (QFT) ?; and (2) more generally, how close can PDE simulations come to replicating the predictions of QFT?

Sections 2 and 3 of this paper will review previous work on the P mapping widely used in quantum optics [2-5], and its extension to classical Hamiltonian field theory in general. For our purposes here, the key result from that work is the generalized operator trace theorem, which, when applied to the function E, yields:

$$\text{Tr}(H_n \rho(S)) = E(S) \qquad (1)$$

where S is the state of the classical system (i.e. a set of values for the fields $\underline{\varphi}(\underline{x})$ and their duals $\underline{\pi}(\underline{x})$ over all points $\underline{x}$ in $R^3$), where $H_n$ is the usual normal form Hamiltonian, where E(s) is the classical Hamiltonian energy of the state S, and where ρ(S) is the density matrix which corresponds to S under the P mapping. From this it easily follows that:

$$\text{Tr}(H_n \rho) = \langle E(S) \rangle , \qquad (2)$$

---

[1] The views expressed here are those of the author, not those of his employer; however, as work produced on government time, it is in the "government public domain." This allows unlimited reproduction, subject to a legal requirement to keep the document together, including this footnote and authorship. Some related material is posted at www.werbos.com.

in the case where:

$$\rho = \int \Pr(S)\rho(S)d^\infty S \tag{3}$$

and where the angle brackets in (2) refer to the expectation value, for any stochastic ensemble of classical states S.

The operator trace theorem clearly tells us that any energy level available in ensembles of the classical system is also present in the spectrum of energy levels allowed in the corresponding QFT, *if* the energy levels are defined by the normal form Hamiltonian $H_n$. However, the classic text on solitons and instantons by Rajaraman [6] states that the lowest energy level available in bosonic QFT which give rise to solitons equals the classical mass-energy *plus* positive correction terms due to stochastic effects; that would imply that the classical soliton mass-energy is lower than the quantum spectrum, and not contained within in.

One possible way to explain this paradox is to note that different versions of QFT are being assumed here. In our work, we are relying on the canonical or Copenhagen version of QFT [7,8], updated to reflect the widespread use of density matrices ρ rather than wave functions ψ in representing stochastic states, as is standard in empirical applied quantum electrodynamics (QED) [4,5,9]. We refer to this version as "KQFT," with "K" for Kopenhagen. (We use "K" rather than "C" because "C" for "cavity" or "circuit" in CQED is already taken, a standard term in important parts of applied QED.) By contrast, Rajaraman's discussion assumes the Feynman path version of QFT, FQFT.

Weinberg [10] reviews the history of how some branches of physics moved from reliance on KQFT to FQFT, for reasons related to the ease of proving certain renormalization results, rather than any kind of decisive empirical test. It is commonly assumed that KQFT and FQFT yield the same predictions, especially for scattering (as in [10]), but in KQFT the point of departure in making predictions is always the normal form Hamiltonian, $H_n$. (See section 6.3 of [8], and [11] for a nice example.) Thus in FQFT, it is generally assumed that the zero point energy or Casimir terms, which are simply deleted when defining $H_n$, are actually present in nature. The classic achievements of KQFT, in explaining anomalous magnetic moments and the Lamb shift, were based on the use of $H_n$, without the need to assume such zero-point energy terms. When zero point energy terms and other such stochastic terms are assumed, it naturally increases predicted masses, except when they are taken back away through renormalization. KQFT correctly predicts the usual flat-plate Casimir experiments on the basis of Vanderwaals forces; however, there are many other experiments which could be discussed, and the choice between KQFT and FQFT in such areas is beyond the scope of this paper.

This paper was motivated in part by the goal of finding a different explanation for the paradox. Could it be that the results for statistical ensembles might be misleading? After all, ensemble states could be mixtures of the vacuum state (zero energy) and higher energy ground states.

The main result of this paper, in section 4, is that we still have a classical-quantum equivalence for the spectra of states of definite energy, but that there is indeed a gap between classical systems and bosonic KQFT which we can quantify. More precisely, we find that any ensemble of classical states S sharing the same definite energy E can be written as:

$$\rho = \int_{S \in E} \Pr(S)\rho(S)d^\infty S = \int c_\alpha |\psi_\alpha\rangle\langle\psi_\alpha| d^?\alpha \quad , \tag{4}$$

where the dimensionality of $\alpha$ depends on the specific PDE system, and where $|\psi_\alpha\rangle$ are the eigenvectors of eigenvalue zero of the fundamental spectral operator:

$$M(E) = N(H_n - E, H_n - E) \quad , \tag{5}$$

where N refers to the normal product. In KQFT, we normally assume that the underlying states of definite energy are eigenvectors of $H_n$, which (because of the nonnegativity of $H_n$) are the same as the eigenvectors of eigenvalue zero of:

$$(H_n - E)(H_n - E) \tag{6}$$

The difference between equation 5 and equation 6 is an operator:

$$\Delta(H_n) = N(H_n, H_n) - H_n H_n \tag{7}$$

To prepare for the results of section 3, section 2 will review some key pieces from the vast literature on the Glauber-Sudarshan P mapping, in the case where the classical system to be quantized is not a field theory but a simple function of two real scalar variables p and q, like the harmonic oscillator. Glauber received the Nobel prize for this important work in this area, which, among other things, played a key role in the development of the laser [3]. Here I will rely heavily on two sources: (1) the definitive more recent book by Carmichael [4], who provides numerous theorems, connections to empirical results, and references to earlier literature; and (2) the key paper of Mehta and Sudarshan [2] which proved the operator trace theorem. Mehta [12] provides a elegant brief recap of the history of the P mapping.

Section 3 briefly shows how to extend the definitions and the operator trace theorem from the case of two scalar variables p and q to the case of two real mathematical vector fields $\underline{\varphi}(\underline{x})$ and $\underline{\pi}(\underline{x})$, as in Hamiltonian field theories. (Mathematical vector fields $\underline{\varphi}$ can actually consist of an amalgam of relativistic covariant vector and tensor fields; thus this applies to a very general class of PDE systems.) It seems likely that the results could be extended still further, to show that spatially localized classical states S map into density operators which are localized around the center of mass, as in the usual quantum mechanical representation of an atom with separation of coordinates; however, that is not proved here.

It should be stressed that this classical-quantum equivalence in expectation values does not imply equivalence in dynamics. As one would expect from the work on Bell's Theorem [9], we have found that the master equations which describe the classical dynamics are quite different in the general case from the usual Schrodinger equation [13].

Section 4 briefly proves that the fundamental spectral operator given in equation 5 has the property claimed. Section 5 discusses some open questions for future research related to these results.

# 2. The P Mapping for a Simple (ODE) Systems

## 2.1. General Concepts

The P mapping provides a 1-to-1 mapping between definite states of the classical system and a subset of the density matrices for the corresponding classical system; it may be viewed either as mapping from classical states to quantum states, or as a mapping from a subset of the quantum states to classical states. Equivalently, it provides a 1-to-mapping between statistical ensembles of classical states and a (larger) subset of density matrices. Carmichael cites prior theorems by Sudarshan showing that any (bosonic) density matrix allowed in QFT can be represented as:

$$\rho = \int P(\alpha) |\alpha\rangle\langle\alpha^*| \, d\alpha \tag{8}$$

where $|\alpha\rangle\langle\alpha^*|$ is the density matrix corresponding to the classical state $\alpha$ under the P mapping, and where P is a real function; however, for some density matrices $\rho$ allowed in QFT, $P(\alpha)<0$ for some states $\alpha$. Carmichael calls such density matrices "nonclassical states." Here we write the operator trace theorem (stated as equation 1.17 of [2]) as:

$$tr(\rho G_n) = \langle g(\alpha) \rangle \tag{9}$$

where g is some function which can be expressed as a polynomial in $\alpha$ and $\alpha^*$, where $\rho$ is the density matrix defined by equation 1, where the classical expected value ($\langle \ \rangle$) is calculated based on $P(\alpha)$ as a probability distribution, and where $G_n$ is the operator which results from quantizing the function f as a normal product, exactly as in KQFT.

## 2.2 Basic Equations of the P Mapping in the ODE Case

Carmichael [4] shows that we can get remarkably far in understanding complex empirical phenomena like decay in two-level atoms and resonance fluorescence by applying the P mapping to simple classical systems like the general harmonic oscillator:

$$H(p,q) = (p^2/2m) + (1/2)m\omega^2 q^2 + H_I(p,q) \tag{10}$$

Carmichael's detailed discussion of the empirical details shows us that the process of an electron dropping down from one energy level to a lower energy level is far more complicated than the simplified story given in first year texts on quantum mechanics.

The classical states of this system are characterized by a complex variable which he defines on page 73, following Glauber's notation:

$$\alpha = (m\omega q + ip)/(2\hbar m\omega)^{-1/2} \tag{11}$$

On page 75, he defines the coherent quantum states in terms of the wave function $|\alpha\rangle$:

$$a|\alpha\rangle = \alpha |\alpha\rangle \quad (12)$$
$$\langle\alpha| a^H = \alpha^* \langle\alpha| \quad (13)$$

where "a" is the usual annihilation operator and where its Hermitian conjugate, $a^H$, is the usual creation operator. From these definitions, he deduces (page 6) that the wave function $|\alpha\rangle$ obeys (and is also defined by):

$$|\alpha\rangle = e^{-\frac{1}{2}|\alpha|^2} \sum_{n=0}^{\infty} \frac{a^n}{\sqrt{n!}} |n\rangle \quad (14)$$

and also (on page 78) that it obeys and is defined by:

$$|\alpha\rangle = e^{-\frac{1}{2}|\alpha|^2} e^{\alpha a^H} |0\rangle \quad (15)$$

Of course, $|\alpha\rangle\langle\alpha^*|$ is the density matrix $\rho$ which the P mapping maps the classical state $\alpha$ into, and it obeys equation 9.

In actuality, the P mapping also works for a broader class of systems than the Hamiltonian systems considered in this paper. Carmichael uses the P mapping to map *from* operator master equations (Schrodinger equations modified so as to approximate the effects of energy dissipating out into a reservoir) *to* sets of classical PDE – Fokker-Planck equations, which provide exact information about the level of dissipation predicted by the master equations. He also uses an extended version of equation 9, based on characteristic functions, so as to calculate two-time averages. He relies heavily on the characteristic function approach, used by Mehta and Sudarshan to establish the fact that $P(\alpha)$ is a well-defined function for all quantum density matrices $\rho$ [2]; however, when we use the mapping in the other direction, from classical ensembles to density matrices $\rho$, it is enough to focus on the case of classical ensembles for which $\Pr(\alpha)$ is a well-defined function. Mehta and Sudarshan use the same relationships, using "z" to represent what we (following Glauber) will call "$\alpha$".

Given any polynomial or analytic function of two variables, p and q, Mehta and Sudarshan [2] note that the function may be represented equivalently as a polynomial function of $\alpha$ and $\alpha^*$, where $\alpha=p+iq$. And so, if we write:

$$g(\alpha, \alpha^*) = \sum_{k,j} A_j^k \alpha^j (\alpha^*)^k \quad (16)$$

Substituting from equations 12 and 13, equation 16 implies:

$$g(\alpha, \alpha^*)|\alpha\rangle\langle\alpha| = \sum_{k,j} A_j^k a^j |\alpha\rangle\langle\alpha| (a^H)^k \quad (17)$$

If we quantize g by quantizing $\alpha$ as the operator a and $a^*$ as $a^H$, the raw quantized version of $g(\alpha,\alpha^*)$ would be:

$$G_r = \sum_{k,j} A_j^k a^j (a^H)^k \quad (18)$$

From equation 17, for |α><α| of trace 1 (a consequence of equations 14 and 15), and using a well-known trace identity, we can deduce:

$$g(\alpha,\alpha^*) = Tr(g(\alpha,\alpha^*)|\alpha\rangle\langle\alpha|) = Tr(\sum_{k,j} A_j^k (a^H)^k a^j |\alpha\rangle\langle\alpha|) \qquad (19)$$

Equation 9 then falls out directly by using a basic property of normal products of expressions such as equations 16 and 18:

$$G_n = \sum_{k,j} A_j^k (a^H)^k a^j \qquad (20)$$

## 3. Extended P Mapping To General Hamiltonian Field Theories

### 3.1. Review of the Extension of the P Mapping to Hamiltonian Field Theories

Here we consider the more general situation where we replace p and q by the real mathematical vector fields $\underline{\varphi}(\underline{x})$ and $\underline{\pi}(\underline{x})$ defined over $\underline{x}$ in $R^3$. A state S of this system is simply a set of values for $\underline{\varphi}(\underline{x})$ and $\underline{\pi}(\underline{x})$ across all $\underline{x}$ in $R^3$. In place of the Hamiltonian shown in equation 9, we assume (as in [13]) a classical Hamiltonian of the form:

$$\mathcal{H} = H = \int \left( \tfrac{1}{2} \sum_{j=1}^{n} (|\nabla \phi_j|^2 + m_j^2 \phi_j^2) + f(\underline{\phi}, \underline{\pi}, \nabla \underline{\phi}) \right) d^3 \underline{x} \qquad (21)$$

Mehta and Sudarshan [2] refer to prior work establishing similar relationships for multiple modes or state variables, $\alpha_j$, and for continuous state variables, which would include state variables indexed by the momentum coordinate $\underline{p}$ in $R^3$. In our case, we replace equation 15 by:

$$|S\rangle = Z e^{\sum_j \int \alpha_j(\underline{p}) a_j^H(\underline{p}) d^3 \underline{p}} |0\rangle, \qquad (22)$$

where Z is the real scalar which normalizes |S> to length 1. Because $a_k(\underline{p}')$ commutes with all the terms in the exponent of equation 14, except for the one where j=k and $\underline{p}'=\underline{p}$, it is easy to see that equation 12 generalizes to:

$$a_k(\underline{p}) |S\rangle = \alpha_k(\underline{p}) |S\rangle \qquad (23)$$

and to its Hermitian conjugate:

$$\langle S| a_k^H(\underline{p}) = \alpha_k^*(\underline{p}) \langle S| \qquad (24)$$

The Hamiltonian H in equation 21 is a fairly complicated polynomial in the momentum representation, in the set of state variables $\varphi_j(p)$ and $\pi_j(p)$ across all $\underline{p}$, involving integrals over $\underline{p}$ (due to Fourier convolution) and appearance of $\underline{p}$ itself (due to the gradient term); however, it is still a polynomial, like equation 16, but with more terms. The logic of equations 16 through 20 still goes through, giving us the generalized operator trace theorem:

$$\text{Tr}(\rho G_n) = \langle g(S) \rangle \qquad (25)$$

Before applying equation 25 to physical systems, we need to decide how to map the state variables $\alpha_j(\underline{p})$ into physical variables of the system, just as Carmichael did for the systems he considered. By analogy to equation 11, we have proposed [15\3]:

$$\alpha_j(\underline{p}) = \theta_j(\underline{p}) + i\tau_j(\underline{p}) \qquad (26)$$

where:

$$\theta_j(\underline{p}) = \sqrt{w_j(\underline{p})} \int e^{-i\underline{p}\cdot\underline{y}} \varphi_j(\underline{y}) d^d \underline{y} \qquad (27)$$

$$\tau_j(\underline{p}) = \frac{1}{\sqrt{w_j(\underline{p})}} \int e^{-i\underline{p}\cdot\underline{y}} \pi_j(\underline{y}) d^d \underline{y} \qquad (28)$$

$$w_j(\underline{p}) = \sqrt{m_j^2 + |\underline{p}|^2} \qquad (29)$$

The normal form quantization of H (or P) results from substituting $\Phi_j(\underline{x})$ and $\Pi_j(\underline{x})$ for $\varphi_j(\underline{x})$ and $\pi_j(\underline{x})$ in equation 21, and mapping classical multiplication into normal forms, where $\Phi_j$ and $\Pi^j$ are defined precisely as in chapter 7 of Weinberg[10]:

$$\Phi_j(\underline{x}) = \Phi_j^+(\underline{x}) + \Phi_j^-(\underline{x}) \qquad (30)$$

$$\Pi_j(\underline{x}) = \Pi_j^+(\underline{x}) + \Pi_j^-(\underline{x}) \qquad (31)$$

where:

$$\Phi_j^-(\underline{x}) = \left(\Phi_j^+(\underline{x})\right)^H \qquad (32)$$

$$\Pi_j^-(\underline{x}) = \left(\Pi_j^+(\underline{x})\right)^H \qquad (33)$$

$$\Phi_j^+(\underline{x}) = c\int \frac{e^{i\underline{p}\cdot\underline{x}} a_j(\underline{p})}{\sqrt{w_j(\underline{p})}} d^d \underline{p} \qquad (34)$$

$$\Pi_j^+(\underline{x}) = -ic\int \left(\sqrt{w_j(\underline{p})}\right) e^{i\underline{p}\cdot\underline{x}} a_j(\underline{p}) d^d \underline{p} \qquad (35)$$

Inserting these definitions into equation 25, and taking normal products, we arrive at the special case of equation 25 most relevant to our purposes here:

$$Tr(\rho H_n) = \langle H(S) \rangle \qquad , \qquad (36)$$

where $H_n$ is the normal form Hamiltonian quantized with the bosonic field operators given in Weinberg [10], and H(S) is the classical Hamiltonian. Of course, the same goes through for the momentum operator and for functions of energy and momentum

### 4. The Fundamental Spectral Operator For P-Mapped Classical Ensembles

The main result of this paper was already stated in the introduction. The claim is that for any ensembles of states of S of a Hamiltonian field theory, all of which have the same definite classical energy level E, that the density matrix ρ for that ensemble (under the P mapping) is spanned by eigenvectors of eigenvalue zero of the fundamental spectral operator M(E) given in equation 5.
    This result follows very directly from the previous results given above. Choose the function:

$$g(S) = (H(S) - E)^2, \qquad (37)$$

where we now use the function H to represent the classical energy function (classical Hamiltonian), and E is a nonnegative real number. Clearly the energy level equals E for all states in the ensemble if and only if g(S) is zero for all states in the ensemble; by the nonnegativity (and lack of multiple zeroes) of g, this is true if and only if the expected value of g in the ensemble is zero. By equation 25, this is true if and only if $G_n = M(E)$ obeys:

$$Tr(M(E)\rho) = 0 \qquad (38)$$

But note that M(E) is a Hermitian operator, nonnegative over all density matrices ρ. Thus it has an orthogonal eigenvector/eigenspace decomposition, with all eigenvalues zero or positive. Likewise, ρ itself, being Hermitian and positive, has its own decomposition into rank 1 states corresponding to its eigenvectors. Because all of the terms in the resulting expansion of Tr(M(E)ρ) are zero or positive, equation 38 can only be satisfied if all of the eigenvectors of ρ are orthogonal to all of the eigenspaces of M(E), except for the eigenspace of eigenvalue zero. (Any component of ρ in the other eigenspaces would yield a positive term, invalidating equation 38). Thus ρ is made up of rank one terms, all made of vectors in that eigenspace of M(E).
    Note that we could have chosen any other polynomial or analytic function g, which has the property that g(S) = 0 for states of energy E and >0 for states of other energy. $G(S)^2$, for example, has the same property. This would yield other operators with the same basic property as M(E), but more complicated. Also, in applying this concept, one may easily extend it to sets of S restricted to zero total momentum or to some desired gauge or rotation angle, if applicable.
    Also note that we cannot construct in general a new version of the Hamitonian operator by simply patching together the zero-eigenvalue eigenspaces of M(E) for

different values of E, because these eigenspaces are not in general orthogonal to each other.

## 5. Questions for the Future

Several questions emerge from these results.

The first question is: a how could we best apply these new mathematical connections in practical areas, such as quantum optics, simulation modeling, or predicting the emergent properties of classical nonlinear dynamical systems? That is an important question, but it gets into many large and complex areas, beyond the scope of this paper.

The second question is – could there be implications for the formulation of QFT itself? Given that the underlying difference between KQFT and FQFT seems to involve the role of the normal form Hamiltonian H versus the raw Hamiltonian, is it possible that predicting spectra based on M(E) rather than equation 6 could be consistent with empirical reality?

At first, this seems unlikely, but equation 7 is actually very close to part of what we actually do in KQFT; see sections 6.3 and 7.1 of Mandl and Shaw [8]. It is basically just the usual second order contraction, which is what we use to calculate the self-energy of the electron. Section 9.6.1 of Mandl and Shaw reminds us that the first great success of QED was in predicting the anomalous magnetic moment of the electron, by Schwinger in 1948. The correction which he was applied was in fact based on the second-order contraction term, rather than any use of time-independent perturbation theory to revise the estimated eigenvalue. For higher-order corrections, Schwinger has noted [14,15] that we can get consistent and accurate predictions simply by bootstrapping the use of physical mass and second-order connections. For other, more routine calculations of atomic and molecular spectra in applied QED, the self-energy corrections are small compared to what is used in applications, but would presumably be similar. Even for more complicated systems in applied QED, such as predicting energy levels in semiconductors, one of the most successful methods has been the Nonequilibrium Green's Function (NGEF) method, which grew out of Schwinger's approach to self-consistent propagators. In fact, it would be interesting to see how much of all this could be deduced as an exercise in phenomenological modeling, similar to Schwinger's source theory, but with density matrices rather than wave functions as the basis of the bootstrapping.

To extend this kind of spectral modeling to the analysis of unexplained spectral data in the nuclear sector [16] could be very important, but would require discussion of which Lagrangian to use, as well as the literature on bosonization, which is far beyond the scope of this paper. (To get a feeling for the size of the bosonization literature, one may go to Google Scholar, and branch forward from the list of papers which cite [11], one of the original seminal papers in that field.) It should be noted that if a Lagrangian is chosen which contains topological Higgs terms, it becomes necessary to map the fields into a kind of equivalent vacuum-dependent representation before the P mapping, like the $\varphi_0$ subtraction used in electroweak theory today, to ensure L2 integrability. Of course, because of the classical-quantum equivalence, this approach always results in finite mass-energies. In a similar vein, Schwinger noted the finite nature of his self-energy correction methods [14,15]. When PDE simulation is used (as in [1]), it is not necessary to have convergent Taylor series or perturbation expressions in order to calculate key spectral predictions.

## Appendix: Thermodynamic Properties of Classical Hamiltonian Systems

This paper has presented a test for an ensemble of classical states S to be an ensemble of definite (uniform) energy. For a large class of classical nonlinear dynamical systems, ODE or PDE, it is also possible to characterize the invariant equilibrium ensembles in a similar manner. More precisely, we show how to extend the Boltzmann distribution (and the more general class  of equilibrium ensembles of which it is an example) to the class of Hamiltonian systems which we call "statistically incompressible."

For the ODE case, consider the usual Fokker-Planck equation, using "p($\underline{x}$)" to represent the density of probability at point $\underline{x}$ in state space:

$$\dot{p} + (\underline{v} \cdot \nabla p) + p(div\, \underline{v}) = 0 \qquad (39)$$

Here, the state space is just the space of possible values for the two vectors $\underline{\varphi}$ and $\underline{\pi}$ in $R^n$. From equation 39, the uniform measure $d^n\underline{\varphi}d^n\underline{\pi}$ (whose gradient is zero) will be an invariant measure (have the property that p dot will be zero) if:

$$"div \, \underline{v}" = \sum_{i=1}^{n} \frac{\partial \dot{\varphi}_i}{\partial \varphi_i} + \sum_{i=1}^{n} \frac{\partial \dot{\pi}_i}{\partial \pi_i} \tag{40}$$

We will call a Hamiltonian ODE system "statistically incompressible" if its dynamics have this property. Likewise, a Hamiltonian PDE system will be called statistically incompressible whenever the measure $d^\infty S = d^{n\infty}\underline{\varphi}(\underline{x})d^{n\infty}\underline{\pi}(\underline{x})$ is invariant under the dynamics of the PDE.

Given any function $g(E,\underline{I})$, where E is energy and $\underline{I}$ is the set of other conserved quantities of the system, it is obvious that $g(E,\underline{I}) \, d^\infty S$ will also be an invariant measure if the system is statistically incompressible. If this measure meets the requirements for a probability distribution (that it be nonnegative and that its integral equals one), then it represents an equilibrium (ergodic) distribution of states S. The generalized Boltzmann distribution is a function of this form ($g = c \exp(-kE-\underline{c}\cdot\underline{I})$).

In addition to the usual primal representation $\rho$ of any ensemble of states S, as given in equation 3, there is also a dual representation F defined by:
$$\Pr(S) = p(S)d^\infty S$$
$$Tr(F\rho(S)) = p(S) \tag{41}$$

Thus the operator version of the Boltzmann distribution is not the usual primal version of the Boltzmann operator, but the dual representation defined by:

$$F = G_n(E,\underline{I}) \tag{42}$$

where g is the usual Boltzmann function (an analytic function).

A previous paper [17] showed how the Bell's Theorem experiment could be predicted by either of two local causal models – using the discrete mathematics of Markov Random Fields, which provide a kind of equilibrium statistical analysis across space and time. It concluded with the question of how to generalize that type of analysis from the discrete case to the case of continuous variables and fields. It is hoped that the observations in this appendix will help clarify and answer that question. The primal and dual P mapping concepts can be extended to stochastic ODE and PDE as well, at the cost of some complexity, beyond the scope of this paper.